\documentclass{Interspeech2024}

\interspeechcameraready

\title{Exploring Speech Foundation Models for Speaker Diarization in Child-Adult Dyadic Interactions}

\name[affiliation={1}]{Anfeng}{Xu}
\name[affiliation={1}]{Kevin}{Huang}
\name[affiliation={1}]{Tiantian}{Feng}
\name[affiliation={2}]{Lue}{Shen}
\name[affiliation={2}]{Helen}{Tager-Flusberg}
\name[affiliation={1}]{Shrikanth}{Narayanan}

\address{
  $^1$University of Southern California, USA\\
  $^2$Boston University, USA }
\email{anfengxu@usc.edu}


\usepackage{threeparttable}
\usepackage{tikz}
\begin{document}

\maketitle

\begin{abstract}
Speech foundation models, trained on vast datasets, have opened unique opportunities in addressing challenging low-resource speech understanding, such as child speech. In this work, we explore the capabilities of speech foundation models on child-adult speaker diarization. We show that exemplary foundation models can achieve $39.5\%$ and $62.3\%$ relative reductions in Diarization Error Rate and Speaker Confusion Rate, respectively, compared to previous speaker diarization methods. In addition, we benchmark and evaluate the speaker diarization results of the speech foundation models with varying the input audio window size, speaker demographics, and training data ratio. Our results highlight promising pathways for understanding and adopting speech foundation models to facilitate child speech understanding.
\end{abstract}

\section{Introduction}
Designing foundation models of broad scope is an emerging research topic in Artificial Intelligence (AI), notably toward creating machine learning models that can understand, interpret, and comprehend a wide range of human-centered signals, either matching or surpassing human-level accuracy \cite{bommasani2021opportunities}. These models are typically trained on massive amounts of publicly available data from the Internet, employing contrastive, reconstructive, or forecasting objectives to learn meaningful and robust data representations. Recently, there have also been significant developments in speech foundation models, such as Whisper \cite{radford2023robust}, WavLM \cite{chen2022wavlm}, and MMS \cite{pratap2024scaling}. The emergence of these models delivers promising speech recognition performances in diverse contexts, notably in low-resource scenarios. Specifically, the primary focus for this work is on child speech understanding.

\begin{figure}[t]
  \centering
  \includegraphics[width=0.96\linewidth]{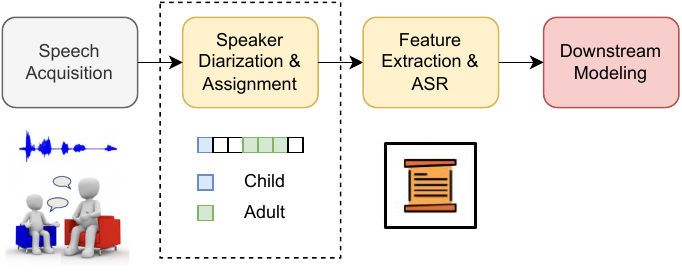}
  \vspace{-3mm}
  \caption{Spoken language assessment pipeline.}
  \label{fig:pipeline}
  \vspace{-3mm}
\end{figure}

Automatic child speech understanding is of paramount importance in developmental clinical contexts, enabling scientists and clinicians to efficiently assess how children interact with others. This process creates numerous opportunities, including neurocognitive disorder assessment or behavioral therapy involving child interactions. Often, these interactions involve dyadic activities with an adult clinician or caregiver, such as child-adult dyadic interaction protocols ADOS-2 \cite{mccrimmon2014test} and ELSA \cite{barokova2021eliciting} for the Autism Spectrum Disorder (ASD) assessment. A fundamental component in analyzing child speech in interaction setting involves accurate and precise speaker diarization, which addresses the question of ``who spoke when" by detecting the speech segments and assigning speaker labels. As shown in Fig.~\ref{fig:pipeline}, speaker diarization is a crucial step towards automatic child speech and behavior analysis, directly impacting the subsequent feature extraction and downstream modeling reliability. For example, speech duration and latency derived from the diarized speech data is a significant biomarker of treatment response in ASD \cite{mckernan2022ASDmarker}.



While recent speaker diarization methods have achieved impressive results on public benchmark datasets, they may not necessarily adapt well to child-adult speaker diarization because of limited child speech data seen in training, especially in light of the significant expected variability and heterogeneity present in child speech and child-involved dialogue. Despite the significant demand for automated child behavior analysis in the clinical domain, challenges persist due to limited availability of child speech data collections and the high costs associated with annotation. Speech foundation models hold great promise in advancing child speech understanding, primarily because of their ability to achieve robust performance in low-resource speech settings. This motivates us to explore speech foundation models for child-adult speaker diarization.


In this work, we propose a child-adult speaker diarization framework that formulates speaker diarization as a frame-level classification problem using speech foundation models. We benchmark speaker diarization performance across nine speech foundation models and provide a comprehensive analysis of how input audio window size, demographics, and data efficiency influence the outcome. To evaluate the effectiveness of the speech foundation models in child-adult speaker diarization, we compare the performance of the speech foundation models with the state-of-the-art (SOTA) speaker diarization systems in the child-adult dyadic setting. The main contributions and findings of the paper are summarized below:

\begin{itemize}[leftmargin=*]

    \item 
    The study provides one of the first attempts exploring speech foundation models for child-adult speaker diarization. 
    
    \item Our results show the effectiveness of integrating foundation models for child-adult speaker diarization, outperforming SOTA speaker diarization methods by offering around $39.5\%$ relative reduction in Diarization Error Rate (DER).

    \item We find that our proposed method using speech foundation model performs robustly across different demographics and achieves strong diarization performance with around 2 hours of training data.
    
\end{itemize}

\section{Background}
\subsection{Speaker diarization}
\label{sec: diarization}
Traditional speaker diarization begins with Voice Activity Detection (VAD) to extract human speech regions, followed by clustering speaker embeddings (e.g., x-vectors \cite{snyder2018x}), which are extracted from these speech regions, to assign speaker labels. More recently, to handle overlapped speech detection, researchers proposed End-to-End Neural Diarization (EEND) \cite{fujita2019end}, which is based on neural network architectures that formulate the speaker diarization task as a classification problem. To improve EEND, researchers have proposed a hybrid model, called EEND-VC \cite{kinoshita2021integrating}, that establishes the diarization as a classification problem on speech sub-segments and subsequently clusters the sub-segment results for the final speaker diarization. Building upon EEND-VC, researchers \cite{kinoshita2021advances, plaquet2023powerset} have achieved competitive results in popular diarization testbeds. In this work, we use VBx \cite{landini2022bayesian} and PyAnnote diarization \cite{Bredin23, plaquet2023powerset} as the clustering and EEND-VC speaker diarization baselines, respectively.



\subsection{Child-adult speaker classification and diarization}
Deep learning approaches have gained attention in utterance-level child-adult speaker classification, delivering improved classification performance \cite{lahiri2023robust, xu2024audio}. However, these works may encounter challenges in real-world deployment as they fail to consider VAD and speaker changes in the model. Conversely, there is limited previous work on child-adult speaker diarization, likely due to the lack of available annotated datasets for this task. While \cite{kumar2020improving, krishnamachari2021developing} are previous works on child-adult speaker diarization using x-vector and clustering, they do not incorporate VAD in their experiments or omit False Alarm when reporting DER. Moreover, \cite{kothalkar2024child} uses CNN for child-adult speaker diarization, but the pipeline requires re-segmentation from an ASR model. \cite{li2023towards} investigates wav2vec 2.0 pre-training with family audio for downstream tasks including frame-level speaker diarization, but DER or comparison to previous methods are not provided and their focus is on infants.


\section{Method}
We use speech foundation models to tackle the child-adult speaker diarization task by formulating it as a frame-level 
classification task. Given the raw input audio $\mathbf{X}$, our goal is to predict the speaker labels $y_t \in \{c, a, o, s\}$ for each audio frame $x_t$, where $t=1,2,\dots,T$ represents the time step such that $[x_1, x_2, ..., x_T] = \mathbf{X}$. The labels $c$, $a$, $o$, and $s$ represent child, adult, overlapped speech, and silence/background noise, respectively. $T$ is the total amount of speech divided by the step size of specific models. All the speech foundation models investigated in this study have a common step size of $20$ ms. 
\subsection{Speech Foundation Models}

\noindent \textbf{Wav2vec 2.0 \& MMS}:
A popular foundation speech model is Wav2vec 2.0 (W2V2) \cite{baevski2020wav2vec}, trained using self-supervised learning (SSL) to predict masked segments of quantized speech representations. Recently, in the MMS project \cite{pratap2024scaling}, researchers pre-trained Wav2vec 2.0 with 1406 languages from around 500k hours of speech data, achieving strong speech recognition results with extensive multilingual capability. We use the base and large models trained on 960 hours of Librispeech \cite{panayotov2015librispeech} for Wav2vec 2.0 and we use the MMS-300m model with approximately 315 million parameters.

\noindent \textbf{WavLM}: WavLM (WLM) \cite{chen2022wavlm} is a speech SSL model that uses k-means clustering to quantize speech segments following the work in \cite{hsu2021hubert}, pre-trained to jointly learn masked speech prediction and denoising. This model yields SOTA performance for various speech recognition and understanding benchmarks. We use the WavLM-base+ and WavLM-large models that were pre-trained on 94k hours of audio data.


\noindent \textbf{Whisper}: Whisper (WSP) is a transformer encoder-decoder model trained by weak supervision for ASR-related tasks, using 680k hours of multilingual speech data collected from the Internet. It provides competitive results on multiple ASR benchmark datasets and works robustly in noisy recording conditions. In this work, we only use the encoder in the child-adult diarization experiments, as it provides hidden representations that are more relevant to our goal. We use Whisper-Tiny, Whisper-Small, Whisper-Base, and Whisper-Medium for our experiments.


\subsection{Downstream architecture}
We first take the weighted average of all hidden layers from each pre-trained speech foundation model, where the weights are learnable. The weighted average is passed to three 1D convolutional layers, each with an output channel size of 256, a ReLU activation function, and a dropout probability of 0.2 during the training. It is then followed by a 1D convolutional layer with output size 4 for the final prediction. The input audio window size of 20 seconds is used for all baseline comparisons, but we investigate the effect of the window size in Section~\ref{sec:window_size}.

\section{Experimental setup}

\subsection{Dataset}
\noindent \textbf{Dataset description}: The dataset comprises 73 video sessions, each capturing a unique child-parent dyadic interaction in English, ensuring no child appears in more than one session. All the children are diagnosed with ASD and are minimally verbal. While recordings were made remotely by the clinical personnel on Zoom, the interacting dyads were physically co-present, as detailed in \cite{Butler, xu2023understanding}. 14 files with the presence of a third speaker (e.g., another parent, sibling) were removed from the original 87 files for a fair comparison under 2-speaker child-adult dyadic interactions. Third-speaker speech segments occasionally exist in some of the remaining files, but these regions are removed through manual inspection. ASD experts have assessed children's spoken language capabilities into three levels: pre-verbal communication (LL-1), first words (LL-2), and word combinations (LL-3) using transcripts, following the guideline in \cite{tager2009defining}. The detailed demographic statistics are shown in Table~\ref{tab:demographics}.

\begin{table}[ht!]
  \centering
  \caption{Dataset statistics.}
  \vspace{-2.5mm}
  \footnotesize
  \begin{tabular}{lc}
    \toprule
    \multicolumn{1}{l}{\textbf{Category}} & \textbf{Statistics}  \\
    \midrule
    Age (month) & Range: $49$ - $95$, Mean: $74.4$, Std: $13.2$ \\
    Child Gender &  $55$ males, $18$ females \\
    Count per LL & $32$ (LL-1), $21$ (LL-2), $20$ (LL-3) \\
    Child Speech Duration & Mean: $153(s)$, Std: $100(s)$  \\
    Adult Speech Duration & Mean: $286(s)$, Std: $96(s)$  \\
    \bottomrule
  \end{tabular}
  \vspace{-3.5mm}
\label{tab:demographics}
\end{table}
\vspace{0.5mm}
\noindent \textbf{Dataset annotations}: The audio data was separately annotated into intelligible speech, unintelligible speech, nonverbal vocalization, and singing utterances, with corresponding child-adult labels by \cite{xu2023understanding}. The original annotation also includes ambiguous (overlap) segments, but these segments are re-annotated carefully so that they do not include single-speaker segments. Utterances are further separated into smaller segments if there are any gaps larger than $200$ ms using an energy-based threshold.

\subsection{Speaker diarization baselines}
\subsubsection{ECAPA-TDNN}
ECAPA-TDNN \cite{desplanques2020ecapa} is a speaker embedding model that shows competitive results for speaker recognition and verification tasks. We use the model from SpeechBrain \cite{speechbrain}, trained on voxceleb1 and voxceleb2 \cite{nagrani2017voxceleb}. We add three convolutional layers of output size 256, followed by one convolutional layer with output size 4, to the temporal output of ECAPA-TDNN for frame-level speaker classification. We freeze the pre-trained ECAPA-TDNN weights during the fine-tuning. 


\subsubsection{VBx}
VBx \cite{landini2020bayesian} is a clustering-based speaker diarization that uses the hidden Markov model to refine the speaker embedding and re-segment the speaker regions based on the preliminary clusters obtained from the Agglomerative Hierarchical Clustering on x-vectors. We use the code and pre-trained models provided by the authors. The 16 kHz x-vector model was trained on voxceleb1, voxceleb2, and cn-celeb \cite{nagrani2017voxceleb, fan2020cn}. As this approach requires an additional VAD component, we experiment with VAD using Ground Truth (GT) VAD and Silero VAD \cite{SileroVAD}. We use $F_A=0.3$, $F_B=12$, and $\text{loop}_P=0.2$ as the hyper-parameters for VBx, determined using grid search.


\subsubsection{PyAnnote}
\label{sec:pyannote}
We use PyAnnote speaker diarization 3.1 \cite{plaquet2023powerset, Bredin23}, which performs a frame-level powerset multi-class classification for local diarization on 5-second windows, followed by Agglomerative Hierarchical Clustering with ECAPA-TDNN \cite{desplanques2020ecapa} embedding to combine the local diarizations. We show the diarization results with and without fine-tuning the segmentation model on our dataset. We also experiment with combining VAD from the fine-tuned PyAnnote diarization with VBx. We set the minimum and maximum number of speakers to one and two. In the case of fine-tuning, we use Adam optimizer with a learning rate of $1\mathrm{e}{-3}$ and batch size of 32, and Cross-Entropy Loss. 

\subsection{Training details for foundation models}
\label{sec:training}
For all speech foundation models, we use the pre-trained models available on HuggingFace \cite{wolf2019huggingface}. We perform 5-fold cross-validation, where the split is at the session level so that there is no overlapping speaker among the train and test set. During the training, $25\%$ of the sessions are held out as validation sets. The same split is used for all the experiments, including the baseline. The training samples are drawn with $50\%$ overlap between the adjacent audio windows. For the test, the audio sliding windows are taken without any overlap. We use Adam optimizer with Cross-Entropy Loss, learning rate $\mathrm{5e}{-4}$, and weight decay $1\mathrm{e}{-4}$ for a maximum of 15 epochs. These hyper-parameters are determined empirically. We use a single NVIDIA GeForce GPU 1080 Ti for training, and it takes between 5 to 15 minutes per epoch for different models experimented within this study.

\subsection{Evaluation}
We report the \textbf{diarization error rate (DER)} in percentage using PyAnnote.metrics \cite{pyannote_metrics}, with a $100$ ms forgiveness collar at the reference audio segment boundaries on each side. We choose $100$ ms instead of the traditional $250$ ms because most child audio segments in this dataset are very short. We report DER with and without skipping the overlapped regions. We also report the \textbf{detection error rate (F+M)}, which is the false alarm rate plus missed detection rate, as well as the \textbf{speaker confusion (SC)} rate. These evaluation metrics are calculated by aggregating the results together from all sessions. Pyannote and VBx do not perform speaker role assignment unlike the other approaches that directly assign child or adult labels, but the detected speakers are optimally matched to child or adult labels using the Hungarian algorithm.

\section{Results \& Discussion}
\subsection{Speech foundation model results}
The speech foundation model diarization results are presented in Table~\ref{tab:main_results}, along with the number of parameters (Size) and pre-training audio size (in Hours). The results show that the size of the models does not substantially impact the diarization outcome with model families in Wav2Vec 2.0 and WavLM. In contrast, for Whisper models, larger models yield substantially lower DERs. We also note that the detection error rate (F+M) is similar across different speech foundation models. Overall, we see that Whisper models perform better than other models. One plausible reason is that Whisper is trained on much more audio data than other models, as we can see that MMS performs substantially better than W2V2-Large because of more available data used in pre-training. The comparisons indicate that child-adult speaker diarization performance increases when the foundation model is pre-trained with more data. We use Whisper-Small for later analysis because of its competitive performance with a relatively small parameter size.

\begin{table}[ht!]
\footnotesize
  \caption{Comparison of speech foundation models in child-adult speaker diarization. F+M and SC indicate detection error rate and speaker confusion, respectively.}
  \label{tab:main_results}
  \centering
  \vspace{-2.5mm}
  \begin{tabular*}{\linewidth}{l c c c c c c}
    \toprule
    \multicolumn{1}{l}{\textbf{Model}} &\textbf{Size} &\textbf{Hours} &\textbf{F+M} & \textbf{SC} & \textbf{DER (*)} \\
    \cmidrule(lr){1-1} \cmidrule(lr){2-3} \cmidrule(lr){4-6} \cmidrule(lr){7-7} 
    W2V2-Base & $95M$ & $960$ & $11.9$ & $13.5$ & $25.3 (23.5)$\\
    W2V2-Large & $315M$ & $960$ & $11.8$ & $12.3$ & $24.1 (22.2)$ \\
    MMS & $315M$ & $500k$ & $11.0$ & $8.4$ & $19.4 (18.1)$ \\
    \cmidrule(lr){1-1} \cmidrule(lr){2-3} \cmidrule(lr){4-6} \cmidrule(lr){7-7} 
    WLM-Base+ & $95M$ & $94k$ & $11.4$ & $11.1$ & $22.5 (20.8)$ \\
    WLM-Large & $315M$ & $94k$ & $11.2$ & $11.1$ & $22.3 (20.5)$ \\
    \cmidrule(lr){1-1} \cmidrule(lr){2-3} \cmidrule(lr){4-6} \cmidrule(lr){7-7} 
    WSP-Tiny & $8.2M$ & $680k$ & $11.3$ & $12.7$ & $24.0 (22.3)$ \\
    WSP-Base & $21M$ & $680k$ & $10.9$  & $9.2$ & $20.1 (18.3)$\\
    WSP-Small & $88M$ & $680k$ & $10.8$ & $6.0$ & $16.7 (14.9)$ \\
    WSP-Medium & $307M$ & $680k$ & $\textbf{10.4}$ & $\textbf{5.2}$ & $\textbf{15.6 (13.9)}$ \\
    \bottomrule
  \end{tabular*}
\begin{tablenotes}\footnotesize
\item * DER without overlap
\end{tablenotes}
\vspace{-2.5mm}
\end{table}

\subsection{Comparing foundation models against the baseline}
To evaluate the effectiveness of child-adult speaker diarization using speech foundation models, we compare them with the baseline approaches as shown in Table~\ref{tab:baseline}. The fine-tuned PyAnnote VAD with VBx achieves the best baseline diarization result with $25.8\%$ DER. Silero VAD + VBx performs the best in the SC metric among the baseline, but this is partially caused by a high Missed Detection rate. It is worth noting that the F+M metric for fine-tuned PyAnnote increases when combined with VBx, likely due to its failure to detect overlapped speech. All speech foundation models perform better than any baseline in terms of DER. Overall, the best child-adult speaker diarization outcome comes from Whisper-Medium, which results in around $39.5\%$ reduction in DER and $62.3\%$ reduction in SC compared to the best baseline approach (fine-tuned PyAnnote VAD + VBx). We also highlight that, in practice, PyAnnote diarization and VBx require longer inference time due to the clustering algorithm. 

\begin{table}[ht!]
\footnotesize
  \caption{Comparison of baseline methods in child-adult speaker diarization. F+M and SC indicate detection error rate and speaker confusion, respectively.}
  \vspace{-2.5mm}
  \label{tab:baseline}
  \centering
  \begin{tabular*}{0.9\linewidth}{l c c c}
    \toprule
    \multicolumn{1}{l}{\textbf{Method}}  &\textbf{F+M} & \textbf{SC} & \textbf{DER (*)}\\
    \cmidrule(lr){1-1} \cmidrule(lr){2-4} 
    ECAPA-TDNN & $15.4$ & $14.6$ & $30.0 (28.3)$ \\
    \cmidrule(lr){1-1} \cmidrule(lr){2-4} 
    GT VAD + VBx & $N.A.$ & $16.3$ & $N.A.$ \\
    Silero VAD + VBx & $22.1$ & $\textbf{11.7}$ & $33.8 (32.0)$ \\
    \cmidrule(lr){1-1} \cmidrule(lr){2-4} 
    PyAnnote & $18.8$ & $16.1$ & $34.9 (34.6)$ \\
    PyAnnote** & $\textbf{10.9}$ & $16.3$ & $27.2 (26.4)$ \\
    PyAnnote VAD** + VBx & $12.0$ & $13.8$ & $\textbf{25.8 (23.7)}$ \\    
    
    \bottomrule
  \end{tabular*}
  
\begin{tablenotes}\footnotesize
\item * DER without overlap
\item ** fine-tuned
\end{tablenotes}
\vspace{-3.5mm}
\end{table}

\subsection{Window size on child-adult speaker diarization}
\label{sec:window_size}
An important factor that may impact the child-adult speaker diarization is the input audio window size. We are particularly interested in investigating whether the audio window size can impact child-adult diarization that relies on speech foundation models. We hypothesize that the larger audio window provides more context for the speech foundation models to disambiguate child and adult speech segments. In Table~\ref{tab:window}, we show the results from varying the audio window size between $5s$, $10s$, $15s$, and $20s$ using the Whisper-Small model. The comparisons indicate that the larger audio window reduces DER, notably in the lower Speaker Confusion rate. One possible explanation is that both child and adult speech are more likely to be seen from a larger audio window. We want to highlight that choosing the appropriate window size depends on the specific application, and for online or streaming applications, it is preferred to rely on a smaller window size that yields faster inference time and shorter latency, compromising though diarization accuracy.

\begin{table}[ht!]
\footnotesize
  \caption{Comparisons of child-adult speaker diarization performances by varying audio window size. The foundation model used in this experiment is Whisper-Small.}
  \vspace{-2.5mm}
  \label{tab:window}
  \centering
  \begin{tabular}{l c c c c}
    \toprule
    \multicolumn{1}{l}{\textbf{Metric}}  &\textbf{5s}   & \textbf{10s} & \textbf{15s} & \textbf{20s}\\
    \cmidrule(lr){1-1} \cmidrule(lr){2-5} 
    $\%$ both child and adult seen & $45.5$ & $64.2$ & $73.7$ & $79.6$ \\
    F+M (Detection Error Rate) & $11.1$ & $10.7$ & $10.7$ & $10.8$ \\
    SC (Speaker Confusion) & $7.7$ & $6.7$ & $6.3$ & $6.0$ \\
    DER (Diarization Error Rate) & $18.8$ & $17.4$ & $17.0$ & $16.7$ \\
    \bottomrule
  \end{tabular}
  \vspace{-3mm}
\end{table}

\subsection{Demographics on child-adult speaker diazization}
\begin{figure}[ht!] {
    \centering
    
    \begin{tikzpicture}
        \node[draw=none,fill=none] at (0,0){\includegraphics[width=0.4\linewidth]{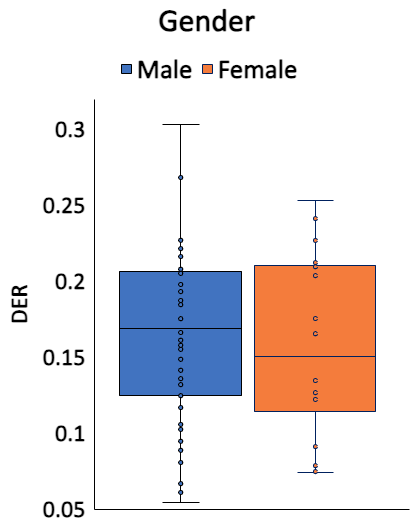}};
        
        \node[draw=none,fill=none] at (0.5\linewidth,0){\includegraphics[width=0.4\linewidth]{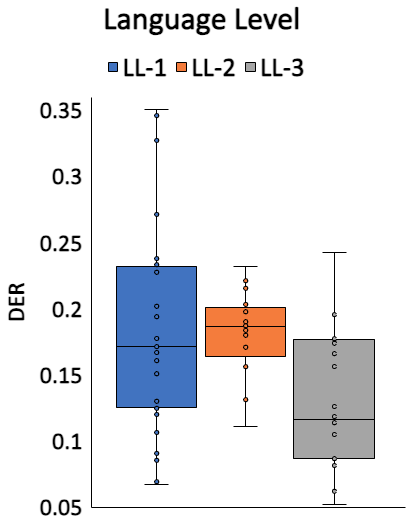}};

    \end{tikzpicture}
    \vspace{-3mm}
    \caption{Comparisons of DER among different demographics (Gender and Language Level). The foundation model used in this experiment is Whisper-Small.}
    \label{fig:boxplots}
    \vspace{-4mm}
    
} \end{figure}

Apart from window size, the demographics of the speakers may play a critical role in diarization performance. To address this question, we present the DER distributions among different gender and language levels, shown in Figure~\ref{fig:boxplots}. The p-value from the Student's t-test of the DER between males and females is 0.304, and the p-value from the ANOVA test of DER from LL-1, LL-2, and LL-3 is 0.045. We can observe that the DER is slightly lower when children have higher language capability (LL-3), but we identify that the diarization performances are overall similar across different demographics. 

\subsection{Data efficiency of speech foundation models}

\begin{figure}[ht!]
  \centering
  \vspace{-4mm}
  \includegraphics[width=0.9\linewidth,height=4.4cm]{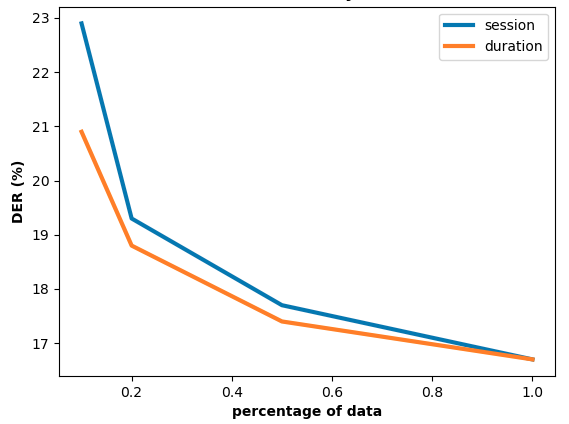}
  \vspace{-3mm}
  \caption{Comparisons of DER at different training data ratios. The foundation model used in this experiment is Whisper-Small.}
  \label{fig:data}
  \vspace{-5mm}
\end{figure}

One major advantage of foundation models is their generalizability to downstream tasks. This motivates us to study whether foundation models can be data efficient in child-adult speaker diarization. This is especially important to academic researchers, given that data annotation is frequently expensive and involves significant time in human training and monetary investments. Thus, it is important to investigate the trade-offs between the amount of annotation effort and the resulting child-adult speaker diarization performance. Here, we explore how reducing the number of sessions or speech duration (e.g., the first 3 minutes instead of the whole 15 minutes) during training affects DER. We vary the training data percentage in both scenarios from $10\%$, $20\%$, $50\%$, and $100\%$, while the validation and test sets remain the same. Figure~\ref{fig:data} shows that with just $20\%$ of training data, which is approximately 130 minutes in our dataset, the diarization based on Whisper-Small achieves below $20\%$ DER for both data downsizing strategies. We also see that decreasing the speech data by session level has a higher impact on DER than reducing the duration.

\section{Conclusion}
The popularity of speech foundation models has opened up unique opportunities in diverse speech recognition and understanding tasks. In this work, we explore speech foundation models for child speech understanding, particularly on child-adult speaker diarization. Our results demonstrate the advantage of incorporating speech foundation models for child-adult speaker diarization, substantially decreasing the diarization error rate compared to SOTA diarization methods. We have also shown that foundation models perform robustly across different demographics and that they achieve strong DER with approximately 2 hours of fine-tuning data. These encouraging results shed light on the importance of further investigating the use of foundation models for broader child speech understanding tasks, including but not limited to automatic speech recognition and multimodal behavior modeling.

\section{Acknowledgements}
This work is supported by funds from Apple.

\bibliographystyle{IEEEtran}
\bibliography{mybib}

\end{document}